\documentstyle[preprint,aps]{revtex}

\begin{document}

\draft

\title{Fibonacci superlattices of narrow-gap III-V semiconductors}

\author{F.\ Dom\'{\i}nguez-Adame, E.\ Maci\'a\thanks{Also at the
Instituto de Estudios Interdisciplinares, El Guijo, Z4 Galapagar,
E-28260 Madrid, Spain.}, and B. M\'endez}

\address{Departamento de F\'{\i}sica de Materiales,
Facultad de F\'{\i}sicas, Universidad Complutense,
E-28040 Madrid, Spain}

\author{C.\ L.\ Roy and Arif Khan}

\address{Department of Physics, Indian Institute of Technology,
Kharagpur 721302, India}

\date{\today}

\maketitle

\begin{abstract}

We report theoretical electronic structure of Fibonacci superlattices of
narrow-gap III-V semiconductors.  Electron dynamics is accurately
described within the envelope-function approximation in a two-band
model.  Quasiperiodicity is introduced by considering two different
III-V semiconductor layers and arranging them according to the Fibonacci
series along the growth direction.  The resulting energy spectrum is
then found by solving exactly the corresponding effective-mass
(Dirac-like) wave equation using tranfer-matrix techniques.  We find
that a self-similar electronic spectrum can be seen in the band
structure.  Electronic transport properties of samples are also studied
and related to the degree of spatial localization of electronic
envelope-functions via Landauer resistance and Lyapunov coefficient.  As
a working example, we consider type II InAs/GaSb superlattices and
discuss in detail our results in this system.

\end{abstract}

\narrowtext

\section{Introduction}

Heterostructures and superlattices (SL) consisting of semiconductors
have been investigated as a source of novel physical properties as well
as for their applications in devices.  From the very beginning, most
researchers have considered the Fibonacci sequence as a typical example
of a quasiperiodic system \cite{Koh,Ostlund} and, some years ago, due to
the advances achieved in nanotechnology, mainly those techniques based
on molecular beam epitaxy, it was possible to fabricate quasiperiodic
semiconductor SL \cite{Merlin,Todd}.  Since these progresses were made,
there has been an increasing interest in the study of (quasiperiodic)
Fibonacci systems, their electronic structure and transport properties.
These studies have provided much information and several physical
properties are now well established.  One of the most conspicuous
features is the occurrence of highly fragmented electronic spectra with
a hierarchy of splitting subbands displaying self-similar patterns
\cite{Laruelle,PRE}.  This exotic electronic spectrum strongly
influences electron propagation \cite{Katsumoto,Angel} and dc
conductance through the system, even at finite temperature
\cite{Enrique}.  In addition, electronic wave functions are neither
extended in the Bloch sense, nor exponentially localised, but critical
in Fibonacci lattices \cite{Chakrabarti}.  All these striking features
make Fibonacci lattices good candidates in regard to investigation of
their novel properties from a theoretical point of view as well as their
potential technological applications in new devices.

Electronic properties of quasiperiodic SL have been studied by various
methods, most of them being based on the envelope-function approximation
\cite{Laruelle}, which is known to be quite successful for periodic
structures \cite{Bastard}.  Neglecting the nonparabolicity of the bands,
the electron dynamics is described by a scalar Hamiltonian corresponding
to decoupled bands in the host semiconductors.  The wave equation is a
Schr\"odinger-like equation for a particle of effective mass $m^*$ in a
one-dimensional potential, so that electron dynamics is studied only
with a single envelope-function.  Since the potential through which the
electron moves is usually regarded as piecewise constant (Kroning-Penney
potential), the solution of the wave equation is a superposition of
plane waves in each layer with real or imaginary momentum, corresponding
to travelling or evanescent solutions respectively.  Matching the
solutions at the interfaces, one can obtain the envelope-function in the
SL using, for instance, the transfer-matrix formalism.  Isotropic and
parabolic bands usually work well in some III-V semiconductors, like
GaAs and AlAs.  However, scalar Hamiltonians cannot adequately describe
narrow-gap semiconductors \cite{Bastard} or those SL whose band
modulation is comparable to the magnitude of the gap, as is the case in
sawtooth-doped GaAs \cite{SST}, since coupling of bands and
non-parabolicity effects for such a situation are usually rather strong.
Hence a more realistic band structure is indeed required to properly
analyse electron states in superlattices of narrow-gap semiconductors.
It is known that two-band models as we report here based on a Dirac-like
equation represent narrow-gap III-V semiconductors quite well
\cite{Callaway}.  In this case, two envelope-functions are needed, one
corresponding to the conduction-band ($s$ like) and the other
corresponding to the valence-band ($p$ like).

The need for a general model like the two-band model poses the question
as to whether the peculiar electronic properties obtained so far within
the one-band model (highly fragmented and self-similar spectra and the
allied transport properties) still remain in narrow-gap SL. As far as we
know, this question has not been answered yet.  The main aim of this
work is to show that those {\em distinctive} characteristics also appear
in more complex and realistic models, suggesting strongly that those
features can be regarded as {\em universal} fingerprints of
one-dimensional Fibonacci systems.  As we have already mentioned, the
equation governing the conduction- and valence-band envelope functions
is a Dirac-like equation.  We can find exact solutions via the
transfer-matrix technique in view of the analogy existing between the
two-band models and the relativistic Dirac theory of electrons.
Transport properties of relativistic electrons in (quasiperiodic)
Fibonacci as well as in (aperiodic) Thue-Morse one-dimensional lattice
have been previously considered by the authors \cite{Angel,Roy1}, and
the transfer-matrix formalism for relativistic electrons is well
established (see Refs.~\onlinecite{Larry,Bruce,FDA,Roy3,Roy2} to cite a
few).  The main difference between the two treatments is that, in the
case of two-band models, the gap is also position-dependent and it
enters in the equation of motion as a scalar-like potential, whereas our
previous relativistic treatments only considered electrostatic-like
potentials (the time component of a Lorentz vector).  Keeping this
difference in mind, one can proceed in analogy with the relativistic
treatment.  Transport properties of the SL at zero temperature are
discussed in the context of transmission coefficient and the Landauer
resistance \cite{Landauer,Roy4} and related to the possible critical
nature of the electronic states.  Spatial extension of the
envelope-functions is determined by means of the Lyapunov coefficient,
which is nothing but the inverse of the localization length.  The nature
of the electronic spectrum of our system is analysed by means of
bandwidth-scaling techniques which suggest a underlying singular
continuous character.

\section{The Model}

The system we study in this work is a Fibonacci superlattice (FSL) made
of two kind of layers of narrow-gap III-V semiconductors, hereafter
denoted by $A$- and $B$-layers.  Due to the offset between conduction-
and valence-band at the interfaces, carriers move under the action of
barriers and wells within the effective-mass approximation.  For
simplicity, we neglect band bending in the rest of the paper, so that
the build-in potential is constant in each layer.  This approximation
simplifies our treatment while keeping the qualitative aspects of the
physics involved.  Without loss of generality, we consider that
$B$-layers act as barriers for electrons and that their width $b$ is
the same for all layers.  To generate our FSL we arrange two tiles $a$
and $a'$ ($a$ and $a'$ larger than $b$), which represent the distance
between two consecutive points characterizing the centres of two
consecutive barriers, according to the Fibonacci sequence (see
Ref.~\onlinecite{Roy2} for further details on how to construct the FSL).
$A$-layers are of thickness $a-b$ or $a'-b$ according to this
arrangement.  The number of barriers in the FSL is a Fibonacci number
$F_l$, obtained from the recurrent law $F_l=F_{l-1}+ F_{l-2}$ with
$F_0=F_1=1$.

After describing the way we construct the FSL, we turn to the dynamics
of electrons in this system.  We treat the resulting electronic
structure by means of the effective-mass ${\bf k}\cdot{\bf p}$
approximation.  The electronic wave function is written as a sum of
products of band-edge orbitals with slowly varying envelope-functions,
assuming that the SL potential is also slowly varying.  To proceed, let
$E_{gA}$ and $E_{gB}$ be the gaps of semiconductors $A$ and $B$
respectively, and let us denote the relative offset between gap centres
by $V$ (in what follows we take the centre of the gap in $A$-layers as
the origin of energies; $V$ is then the energy of the gap centre in
$B$-layers which could be positive or negative, depending of the kind of
interfaces).  As pointed out earlier, we will restrict ourselves to the
case of two nearby bands.  Thus, there are two coupled
envelope-functions describing the conduction-band and valence-band
states of the semiconductor, subject to an effective $2\times 2$
Dirac-like equation.  Assuming that both the gap and the gap centre
depend only on $x$ (the growth direction), the resulting equation for
the envelope-functions in the conduction- and valence-band can be
written as
\begin{equation}
\left[ \begin{array}{cc} {1\over 2}E_g(x)-E+V(x) & -i\hbar v \partial \\
-i\hbar v \partial & -\,{1\over 2}E_g(x)-E+V(x) \end{array} \right]
\left( \begin{array}{c} f_c (x)\\ f_v(x)\end{array} \right) =0,
\label{Dirac}
\end{equation}
where $\partial =d/dx$. Here, $E_g(x)$ is the position-dependent gap and
$V(x)$ gives the energy of the gap centre.  The velocity $v$ is related
to the Kane's momentum matrix elements and is given by $v^2=E_g/2m^*$.
In spite of the fact that both $E_g$ and the effective mass $m^*$ are in
general position dependent, the value of $v$ is almost constant in
direct gap III-V semiconductors \cite{Beresford}.  Hereafter, we assume
this constancy in the SL.  It should be mentioned that the non-zero
in-plane momentum can be easily absorbed in the definition of parameters
and we will henceforth ignore it.

Considering only electronic states below the barrier, which are of most
interest to study quantum confinement effects, the solution of the
Eq.~(\ref{Dirac}) yields the following expression with reference to the
{\em n\/}th barrier centered at $x_n$
\begin{mathletters}
\label{solution}
\begin{eqnarray}
\left( \begin{array}{c} f_{cn}^A (x)\\ f_{vn}^A(x)\end{array} \right)
& = & p_n \left( \begin{array}{c} 1\\ \gamma \end{array} \right)
\exp [i \kappa (x-x_n-b/2)]+
q_n \left( \begin{array}{c} 1\\ -\gamma \end{array} \right)
\exp [-i \kappa (x-x_n-b/2)] \nonumber \\
& & x_n+b/2 < x < x_{n+1}-b/2, \label{solutiona}\\
\left( \begin{array}{c} f_{cn}^B (x)\\ f_{vn}^B(x)\end{array} \right)
& = & s_n \left( \begin{array}{c} 1\\ i\lambda \end{array} \right)
\exp (-\eta x)+
u_n \left( \begin{array}{c} 1\\ -i\lambda \end{array} \right)
\exp (\eta x) \nonumber \\
& & x_n-b/2 \leq x \leq x_n+b/2, \label{solutionb}
\end{eqnarray}
\end{mathletters}
where, for brevity, we have defined the following real parameters
\begin{mathletters}
\label{parameters}
\begin{eqnarray}
\kappa & = & \left( {1\over \hbar v}\right) \sqrt{E^2-E_{gA}^2/4},
               \label{parametersa} \\
\gamma & = & {E-E_{gA}/2\over \hbar v \kappa},
               \label{parametersb} \\
\eta & = & \left( {1\over \hbar v}\right) \sqrt{E_{gB}^2/4-(E-V)^2},
               \label{parametersc} \\
\lambda & = & {E_{gB}/2-E+V\over \hbar v \eta}.
               \label{parametersd}
\end{eqnarray}
\end{mathletters}

Assuming the continuity of the envelope-functions at the interfaces, we
can eliminate $(s_n,u_n)$, thus relating $(p_n,q_n)$ with $(p_{n-1},
q_{n-1})$ via the $2\times 2$ transfer-matrix $M(n)$ through the
relationship
\begin{equation}
\left( \begin{array}{c} p_n\\ q_n\end{array} \right) = M(n)
\left( \begin{array}{c} p_{n-1}\\ q_{n-1}\end{array} \right) \equiv
\left( \begin{array}{cc} \alpha_n & \beta_n \\
\beta_n^* & \alpha_n^* \end{array} \right)
\left( \begin{array}{c} p_{n-1}\\ q_{n-1}\end{array} \right),
\label{M}
\end{equation}
where
\begin{mathletters}
\label{elements}
\begin{eqnarray}
\alpha_n & = & \left[ \cosh(\eta b) +i\, \left( {\gamma^2-\lambda^2
\over 2\gamma\lambda} \right) \sinh (\eta b) \right]
\exp [ i\kappa (\Delta x_n -b) ] \label{elementsa} \\
\beta_n & = & -i\, \left( {\gamma^2+\lambda^2
\over 2\gamma\lambda} \right) \sinh (\eta b)\,
\exp [ -i\kappa (\Delta x_n -b) ] \label{elementsb},
\end{eqnarray}
\end{mathletters}
with $\Delta x_n \equiv x_n-x_{n-1}$ and the convention $x_0=0$.
Note that $\mbox{det}[M(n)]=1$. Letting $N$ be the total number of
barriers ($B$-layers), the transfer matrix $T(N)$ of the whole FSL is
obtained as follows
\begin{equation}
T(N)=\prod_{n=N}^1\> M(n) \equiv \left( \begin{array}{cc} A_N & B_N \\
B_N^* & A_N^* \end{array} \right).
\end{equation}
The elements of the transfer matrix $T(N)$ can be easily calculated
recursively taking into account the fact that $T(N)=M(N)\,T(N-1)$. In
particular we find the expression \cite{Angel,PRB2}
\begin{equation}
A_N=\left( \alpha_N + \alpha_{N-1}^*{\beta_N \over \beta_{N-1}} \right)
A_{N-1} -\,\left( {\beta_N \over \beta_{N-1}} \right) A_{N-2},
\end{equation}
supplemented by the initial conditions $A_0=1$, $A_1=\alpha_1$.

Once we have calculated the matrix element $A_N$, some physically
relevant entities can be readily obtained from it.  Thus, the
transmission coefficient $\tau$ at a given energy $E$ is written as
\begin{equation}
\tau={1 \over |A_N|^2}.
\label{tau}
\end{equation}
Also the single-channel, dimensionless Landauer resistance is given as
\cite{Landauer}
\begin{equation}
\rho = {1-\tau\over \tau}=|A_N|^2-1.
\label{rho}
\end{equation}
The dependence of the resistance with the system size is useful to study
the spatial extension of the electronic states: Localized states lead to
a nonhomic behaviour of the resistance which increases exponentially
with the system size, whereas extended states show a bounded resistance.
Apart from these two entities, there are other which can also be
obtained from $A_N$. Indeed, the Lyapunov coefficient $\Gamma$ is a
nonnegative parameter given by \cite{Kirk}
\begin{equation}
\Gamma=-\,{1\over 2N} \ln \tau.
\label{Lyapunov}
\end{equation}
The Lyapunov coefficient represents the growth rate of the
envelope-function and it is nothing but the inverse of the localization
length in units of the SL period: The more localized the electronic
state, the larger the value of $\Gamma$.

Finally, considering periodic boundary conditions at both edges of the
FSL, one can obtain the following condition for an energy to lie in an
allowed miniband
\begin{equation}
\left({1\over 2}\right)\left|\mbox{Tr}\left[ T(N) \right]\right|\leq 1.
\label{periodic}
\end{equation}
In particular, in the case of periodic SL ($a=a'$) with period $L$ the
Bloch theorem holds good and one then gets $\cos KL=\mbox{Re}\,
(\alpha_1)$; for this situation the dispersion relation $E(K)$, $K$
being the crystal momentum, is found from
\begin{equation}
\cos KL = \cosh (\eta b) \cos[\kappa (L-b)] - \left( {\gamma^2-\lambda^2
\over 2\gamma\lambda} \right) \sinh (\eta b)  \sin [\kappa (L-b)].
\label{monobarrier}
\end{equation}
This expression will be used later to determine the miniband structure
in periodic SL, in order to compare it with results obtained in FSL.

\section{I\lowercase{n}A\lowercase{s}/G\lowercase{a}S\lowercase{b}
FSL}

As an application of our results we consider nearly lattice-matched
InAs/GaSb SL. Recently, much interest was centered on the use of these
materials in resonant tunnelling devices which produce differential
negative resistance with high peak-to-valley current ratio even at room
temperature \cite{Soder}.  The band alignment is type II-staggered
\cite{Gau} in the InAs/GaSb interface, as shown in Fig.~\ref{fig1}.
Perhaps this is the most interesting feature because the conduction-band
edge of the InAs is $0.15\,$eV lower in energy than the valence-bad edge
of the GaSb \cite{Esaki}, so that electrons can flow from the
conduction-band of the InAs to the valence-band in GaSb.  Moreover,
these two semiconductors present a nearly equal Kane's matrix element
leading to $\hbar v \sim 0.77\, $eV\,nm, thus supporting our previous
assumption that this parameter is constant through the whole SL. From
Fig.~\ref{fig1} we conclude that $E_{gA}=0.36\,$eV, $E_{gB}=0.67\,$eV,
and $V=0.665\,$eV.  We set layer thickness leading to $a=6.0\,$nm,
$a'=6.2\,$nm, and $b=4.0\,$nm in our numerical computations.

Using the above set of parameters, we first studied the miniband
structure using (\ref{monobarrier}) in periodic SL with periods $L=a$
and $L=a'$.  The corresponding dispersion relations are shown in
Fig.~\ref{fig2}.  The most relevant feature is that, in both cases,
there exists only one miniband between the conduction-band edges of InAs
($0.18\,$eV) and GaSb ($1.0\,$eV).  As expected, the larger the SL
period $L$, the deeper the miniband.  Nevertheless, this is the only
noticeable effect since the miniband-width is almost unchanged

Now we consider the most prominent features of the resulting electronic
structure when quasiperiodicity is introduced ($a\neq a'$).
Figure~\ref{figx} presents a schematic diagram of the band-edge profile
in the FSL. As we have already explained in the Introduction, one of the
most characteristic properties of electronic spectra in (one-band) FSL
is its highly fragmented, Cantor-like nature.  We have confirmed this
fragmentation in our (two-band) FSL, even when deviation from perfect
periodicity is actually small.  This deviation can be quantitatively
measured from the ratio $a'/a$ which, with our choice of parameters, is
very close to unity.  In fact, using the condition (\ref{periodic}), we
have found that the miniband of the periodic SL splits into several
sub-minibands, that is, small gaps appear.  The origin of these small
inner gaps are directly related to the loss of long-range quantum
coherence of the electrons.  Results corresponding to the fragmentation
of the miniband are shown in Fig.~\ref{fig3} as a function of the
Fibonacci order $l$, i.\ e., the number of GaSb layers in the FSL is
$F_l$.  Only short approximants of the FSL are displayed since on
increasing $l$ the spectrum becomes so fragmented that it is difficult
to observe minor features in the plot.  However, we have carefully
analysed FSL spectra up to order $l=15$ ($N=F_{15}=987$ GaSb layers) and
we have confirmed that the number of sub-minibands composing the whole
spectrum is exactly $F_l$.  The two outermost main sub-mibands present
$F_{l-2}$ subsub-minibands whereas the innermost sub-miniband presents
$F_{l-3}$ subsub-minibands so the total number of subsub-miniband is
$F_{l-2} + F_{l-3} + F_{l-2}=F_{l-1}+F_{l-2}=F_l$.  The number of
subsub-minibands in each main cluster is a consequence of how the energy
spectrum is fragmented, as we shall further discuss later.  Since,
strictly speaking, quasiperiodicity is only observable in the limit $N
\rightarrow \infty$, our results provide information on the prefractal
signature of the FSL energy spectrum.  We have observed that both the
position and widths of the main sub-minibands converge very rapidly to
stable values with increasing the number or GaSb barriers.  We shall
refer to this behaviour as {\em asymptotic stability} of the spectrum; it
implies that its global structure can be obtained in practice by
considering very short approximants, as short as $F_9=55$ barriers, to
very large FSL.

Another characteristic feature of Fibonacci systems is the self-similar
pattern exhibited by their corresponding spectra.  This self-similarity
has been widely investigated within the tight-binding approximation and,
to lesser extend, in wide-gap semiconductor FSL.  Our results point
out that self-similar spectra are also obtained in narrow-gap
semiconductor FSL as shown in Fig.~\ref{fig4}.  It is clear that the
whole electronic spectrum for a short approximant ($F_3=3$ in this case)
is mapped onto a small portion of the spectrum of a higher approximants
($F_6=13$ and $F_9=55$ in Fig.~\ref{fig4}).  This is a consequence of
how the FSL is constructed, based on a deterministic substitution
sequence \cite{Kohmoto}.  Our previous experience \cite{Enrique} has led
us to the conclusion that the fragmentation scheme of a particular kind
of Fibonacci lattice is very well characterized by means of the Lyapunov
coefficient.  Hence, we undertook the study of this parameter in InAs
/GaSb FSL. Results are displayed in Fig.~\ref{fig5} for a FSL with
$N=F_{11}=144$ barriers, although results are independent of the
Fibonacci order.  To be specific, in all cases we have considered we
have observed a well-defined trifurcation pattern of the energy
spectrum, characterized by the presence of three main sub-minibands
separated by large minigaps.  Inside each main sub-miniband, the
fragmentation scheme follows a trifurcation pattern in which each
sub-miniband further trifurcates obeying a hierarchy of splitting from
one to three subsub-minibands.  This fragmentation scheme is clearly
observed in Fig.~\ref{fig5}(a), in which the three main sub-minibands
are detected as an overall decrease of the Lyapunov coefficient, whereas
minigaps appear as local maxima.  In Fig.~\ref{fig5}(b), an enlarged
view of the lower main sub-miniband shows the self-similar nature of the
spectrum structure.

In the thermodynamical limit Fibonacci systems present singular
continuous electronic spectrum \cite{Ghez}.  In order to estimate the
spectral type associated with our FSL, we have computed the so-called
equivalent bandwidth $S$, defined as the sum of all allowed
sub-minibands.  As can be expected from the Cantor-like nature of
Fibonaccian spectra, $S$ vanishes as the number of barriers increases
according to a power law of the form $S=F_N^{-\beta}$ with $\beta \sim
0.3$ (see Fig.~\ref{fig6}).  Earlier works \cite{Kohmoto} reported
that such power-law scaling is characteristic of a singular continuous
spectrum for which all the wave functions are critical, i.e., regarding
localization properties, the functions are neither exponentially
localized nor extended in the Bloch sense.  Therefore, it should be
expected that FSL present higher values of the Landauer resistance at
zero temperature than periodic SL with the same number of barriers
since, in the former case, electronic states are critical whereas in the
later case electronic states are truly extended (Bloch states).  To
confirm this situation we have evaluated the Landauer resistance $\rho$
by means of its definition (\ref{rho}) for the two kinds of SL with
$N=F_8=34$ GaSb barriers, as shown in Fig.~\ref{fig7}.  Notice that, in
addition to the occurrence of well-define inner gaps leading to a strong
enhancement of $\rho$, the overall resistance is larger in the case of
FSL. These results suggest that the overall resistance of the SL is
directly connected with the decay rate of the electronic
envelope-function along the sample, as can be deduced from the
comparison between Fig.~\ref{fig5} showing the inverse of the
localization length, and Fig.~\ref{fig7}(b).  Hence, the spatial
extension of envelope-functions controls the electrical transport of the
sample, as is the case for wide-gap semiconductor FSL.

\section{Summary}

We have studied theoretically a new type of quasiperiodic SL made of
narrow-gap III-V compounds, whose electron dynamics is described by
means of an effective Dirac equation in the framework of an
effective-mass ${\bf k}\cdot{\bf p}$ approach.  The quasiperiodic SL is
constructed arranging two kind of narrow-gap semiconductor layers
following the Fibonacci sequence, assuming that barrier thickness is
always the same while well thickness takes on two values according to
the inflation rule of the Fibonacci series.  By means of the
transfer-matrix formalism, we obtain closed expressions to study
electron transport through the Landauer resistance, localization length
of electrons through the Lyapunov coefficient, and the spectral nature
of the FSL. Those expressions are suitable for an efficient numerical
treatment.  Although the method is completely general even in the
presence of band-inverted semiconductors as it is the case of some IV-VI
heterostructures (e.\ g., Pb$_{1-x}$Sn$_{x}$Te), we have focused our
attention on InAs/GaSb FSL. The corresponding electronic spectrum shows
a highly fragmented, self-similar nature resembling that found for
simpler tight-binding models.  The spectral nature of our model
Hamiltonian, obtained from bandwidth-scaling considerations, indicates
that it is singular continuous in the thermodynamical limit, in
agreement with the conjecture that the spectral type for almost all
substitution sequences should be singular continuous \cite{Ghez}.  Using
the Lyapunov coefficient we have been able to demonstrate that the
electronic spectrum follows a trifurcation scheme of fragmentation with
increase of the Fibonacci order.  In addition, investigations on the
Landauer resistance indicate an overall increase of its value as
compared to periodic SL. The relationship obtained in this regard
suggests that the resistance of the FSL is directly connected with the
decay rate of the electron envelope-function along the sample.

\acknowledgments

The authors thank A.\ S\'anchez for a critical reading of the
manuscript.  Work at Madrid is supported by UCM under project
PR161/93-4811.  Arif Khan is grateful to CSIR, India, for awaring him a
senior research fellowship.



\begin{figure}
\caption{Schematic band-edge diagram of InAs/GaSb type II interface,
neglecting band bending at the heterojunction.}
\label{fig1}
\end{figure}

\begin{figure}
\caption{Miniband structures for periodic InAs/GaSb superlattices with
periods $L=6.0\,$nm (solid line) and $L=6.2\,$nm (dashed line) and
barrier thickness $b=4.0\,$nm. Energies are measured from the gap
centre of InAs.}
\label{fig2}
\end{figure}

\begin{figure}
\caption{Schematic band-edge diagram of a InAs/GaSb FSL. GaSb layers are
of the same thickness $b$ and they are centered at $x_n$. $x_n-x_{n-1}$
takes on two values $a$ or $a'$ according to the Fibonacci sequence.}
\label{figx}
\end{figure}

\begin{figure} \caption{Allowed sub-minibands as a function of the
Fibonacci order $l$, for a InAs/GaSb FSL. The number of sub-minibands is
$F_l$ for each order $l$.  Thickness of GaSb layers is $b=4.0\,$nm and
the distance between their centres is $a=6.0\,$nm or $a'=6.2\,$nm,
arranged according to the Fibonacci sequence.}
\label{fig3}
\end{figure}

\begin{figure}
\caption{Self-similar spectrum of the InAs/GaSb FSLs with the same layer
thickness as in Fig.~3.  Left plot shows the whole spectrum of a FSL of
order $l=3$ whereas central and right plots show a detail of the
spectrum of the FSL of order $l=6$ and $l=9$, respectively. }
\label{fig4}
\end{figure}

\begin{figure}
\caption{(a) Lyapunov coefficient as a function of energy for InAs/GaSb
FSL with the same layer thickness as in Fig.~3.  The number of GaSb
layers is $F_{11}=144$.  (b) An enlarged view of one of the main
sub-mibands in which the self-similar character is more clearly
observed.}
\label{fig5}
\end{figure}

\begin{figure}
\caption{Equivalent bandwidth $S$ as a function of the number of GaSb
layers $N=F_l$ for a InAS/GaSb FSL with the same layer thickness as in
Fig.~3.}
\label{fig6}
\end{figure}

\begin{figure}
\caption{Landauer resistance as a function of energy for InAs/GaSb SLs
for (a) periodic with $L=6.0\,$nm and $b=4.0\,$nm (b) Fibonacci with the
same parameters as in Fig.~2.  In both cases the number of GaSb layers
is $F_{8}=34$.}
\label{fig7}
\end{figure}


\begin{references}

\bibitem{Koh} M.\ Kohmoto, L.\ P.\ Kadanoff, and C.\ Tang, Phys.\ Rev.\
Lett.\ {\bf 50}, 1870 (1983).

\bibitem{Ostlund} S.\ Ostlund and R.\ Pandit, Phys.\ Rev. B {\bf 29},
1394 (1984).

\bibitem{Merlin} R.\ Merlin, K.\ Bajema, R.\ Clarke, F.\ -Y. Juang, and
P.\ K.\ Bhatacharya, Phys.\ Rev.\ Lett.\ {\bf 55}, 1768 (1985).

\bibitem{Todd} J.\ Todd, R.\ Merlin, R.\ Clarke, K.\ M.\ Mohanty, and
J.\ D.\ Axe, Phys.\ Rev.\ Lett.\ {\bf 57}, 1157 (1986).

\bibitem{Laruelle} F.\ Laruelle and B.\ Etienne, Phys.\ Rev. B {\bf 37},
4816 (1988).

\bibitem{PRE} E.\ Maci\'a, F.\ Dom\'{\i}nguez-Adame and A.\ S\'anchez,
Phys.\ Rev.\ E {\bf 50}, R679 (1994).

\bibitem{Katsumoto} S.\ Katsumoto, N.\ Sano, and S.\ Kobayashi, Solid
State Commun. {\bf 85}, 223 (1993).

\bibitem{Angel} F.\ Dom\'{\i}nguez-Adame and A.\ S\'anchez, Phys.\
Lett.\ A {\bf 159}, 153 (1991).

\bibitem{Enrique} E.\ Maci\'a, F.\ Dom\'{\i}nguez-Adame, and A.\
S\'anchez, Phys.\ Rev.\ B {\bf 49}, 9503 (1994).

\bibitem{Chakrabarti} A.\ Chakrabarti, S.\ N.\ Karmakar, and R.\ K.\
Moitra, Phys.\ Lett.\ A {\bf 168}, 301 (1992).

\bibitem{Bastard} G.\ Bastard, {\em Wave Mechanics Applied to
Semiconductor Heterostructures} (Editions de Physique, Paris, 1989).

\bibitem{SST} F.\ Dom\'{\i}nguez-Adame and B.\ M\'{e}ndez, Semicon.\
Sci.\ Technol.\ {\bf 9}, 1358 (1994).

\bibitem{Callaway} J.\ Callaway, {\em Quantum Theory of the Solid State}
(Academic Press, New York, 1991) p 36.

\bibitem{Roy1} C.\ L.\ Roy and Arif Khan, J.\ Phys.:\ Condens.\ Matter
{\bf 6}, 4493 (1994).

\bibitem{Larry} M.\ L.\ Glasser and S.\ G.\ Davison, Int.\ J.\ Quantum
Chem.\ {\bf IIIs}, 867 (1970).

\bibitem{Bruce} B.\ H.\ J.\ McKellar and G.\ J.\ Stephenson, Phys.\
Rev.\ C {\bf 35}, 2262 (1987).

\bibitem{FDA} F.\ Dom\'{\i}nguez-Adame, J.\ Phys.:\ Condens.\ Matter
{\bf 1}, 109 (1989).

\bibitem{Roy3} C.\ L.\ Roy and Arif Khan, J.\ Phys.:\ Condens.\ Matter
{\bf 5}, 7701 (1993).

\bibitem{Roy2} C.\ L.\ Roy and Arif Khan, Phys.\ Rev.\ B {\bf 49},
14\,979 (1994).

\bibitem{Landauer} R.\ Landauer, IBM J.\ Res.\ Dev.\ {\bf 1}, 223
(1957).

\bibitem{Roy4} C.\ L.\ Roy and Chandan Basu, Phys.\ Rev.\ B {\bf 45},
14\,293 (1992).

\bibitem{Beresford} R.\ Beresford, Semicond.\ Sci.\ Technol.\ {\bf 8},
1957 (1993).

\bibitem{PRB2} A.\ S\'anchez, E.\ Maci\'a, and F.\ Dom\'{\i}nguez-Adame,
Phys.\ Rev.\ B {\bf 49}, 147 (1994).

\bibitem{Kirk} P.\ D.\ Kirkman and J.\ B.\ Pendry, J.\ Phys.\ C {\bf
17}, 4327 (1984).

\bibitem{Soder} J.\ R.\ Soderstrom, D.\ H.\ Chow, and T.\ C.\ McGill,
Appl.\ Phys.\ Lett.\ {\bf 55}, 1094 (1989).

\bibitem{Gau} G.\ J.\ Gaultieri, G.\ P.\ Schwartz, R.\ G.\ Nuzzo, R.\
J.\ Malik, and J.\ F.\ Walker, J.\ Appl.\ Phys.\ {\bf 61}, 5337 (1987).

\bibitem{Esaki} L.\ L.\ Chang and L.\ Esaki, Surf.\ Sci.\ {\bf 98}, 70
(1980).

\bibitem{Kohmoto} M.\ Kohmoto, Phys.\ Rev.\ Lett.\ {\bf 51},
1198 (1983).

\bibitem{Ghez} A.\ Bovier and J.\ -M.\ Ghez, Commun.\ Math.\ Phys.\
{\bf 158}, 45 (1993).

\end{references}
\end{document}